# LLM-Assisted Tool for Joint Generation of Formulas and Functions in Rule-Based Verification of Map Transformations

Ruidi He
*Institute for Software and Systems Engineering*
*Technische Universität Clausthal*
Clausthal-Zellerfeld, Germany
ruidi.he@tu-clausthal.de

Yu Zhang
*Institute for Software and Systems Engineering*
*Technische Universität Clausthal*
Clausthal-Zellerfeld, Germany
yu.zhang.1@tu-clausthal.de

Meng Zhang
*Institute for Software and Systems Engineering*
*Technische Universität Clausthal*
Clausthal-Zellerfeld, Germany
meng.zhang@tu-clausthal.de

Andreas Rausch
*Institute for Software and Systems Engineering*
*Technische Universität Clausthal*
Clausthal-Zellerfeld, Germany
andreas.rausch@tu-clausthal.de

*Abstract*—High-definition map transformations are essential in autonomous driving systems, enabling interoperability across tools. Ensuring their semantic correctness is challenging, since existing rule-based frameworks rely on manually written formulas and domain-specific functions, limiting scalability.

In this paper, We present an LLM-assisted pipeline that jointly generates logical formulas and corresponding executable predicates within a computational FOL framework, extending the map verifier in CommonRoad scenario designer with elevation support. The pipeline leverages prompt-based LLM generation to produce grammar-compliant rules and predicates that integrate directly into the existing system.

We implemented a prototype and evaluated it on synthetic bridge and slope scenarios. The results indicate reduced manual engineering effort while preserving correctness, demonstrating the feasibility of a scalable, semi-automated human-in-the-loop approach to map-transformation verification.

## I. INTRODUCTION

Autonomous driving systems rely on high-quality digital maps to support perception, planning, and control tasks [1], [2]. These maps are available in different formats, such as OpenDRIVE [3] and CommonRoad [4], each offering a specific geometric and semantic abstraction of the real world. During system development and simulation, it is often necessary to transform maps from one representation to another for compatibility across tools and platforms.

Such transformations go beyond mere syntactic changes: they involve semantic transformations that may require complex computations, potentially introducing information loss or approximation errors [5], [6]. For example, geographic coordinates (latitude/longitude) may be converted to Cartesian coordinates with elevation, or roads may be represented as polylines in one format and as polynomial curves in another. Ensuring that such transformations preserve critical semantic and geometric information is essential for safe autonomous driving.

The correctness of these transformations is crucial, yet there is no simple universal criterion to verify them. While transformation tools could be tested with extensive test suites, such tests are costly to construct and cannot cover the full range of transformation cases due to the complexity of map data [6].

An alternative is to validate the transformation result itself using rule-based verification frameworks, such as the CommonRoad verification framework proposed by Althoff et al. [7]. These systems check for structural and semantic correctness using formalised rules. However, writing such rules manually is time-consuming, error-prone, and inevitably incomplete, especially as new semantic domains are introduced.

Recent work has shown that large language models can generate Boolean logic, propositional logic, and even more expressive systems like predicate logic (first-order logic) [8], [9]. However, producing complex logical structures remains challenging and prone to inaccuracies [10]. Consequently, many existing LLM-based verification approaches target computational first-order logic (FOL), where formulas are expressed symbolically, and the atomic symbols (predicates and functions) are implemented in imperative code and invoked during formula evaluation [11]. This paradigm has proven highly effective in rule-based verification, for example in the context of SMT solving (e.g., SMT-LIB) [12].

Yet, when verification must be extended to new semantic domains, such as validating complex map transformations. it is insufficient merely to generate new FOL formulas. In these cases, the LLM must also propose the corresponding functions that can be invoked within formulas to access and process domain-specific information [9]. In effect, formulas and functions are inherently interdependent, and both must be generated in tandem to form an executable ruleset.

To the best of our knowledge, no existing research has explored using LLM for the joint generation of logical formulas and corresponding functions in computational FOL to enable scalable verification across new semantic domains. In particular, combining LLM-generated formulas with semantically defined function implementations within an integrated framework for map transformation validation remains an open and unexplored research direction [5], [7].

**Research Challenge:** The central challenge of this work is to investigate whether large language models (LLMs) can be leveraged to generate both formulas and their corresponding functions for use in computational first-order logic (FOL) within the context of rule-based verification of map transformations. Unlike prior approaches that only target symbolic rule generation, this requires the LLMs to co-generate logical rules and the executable functions they depend on, ensuring that the extended ruleset is directly integrable into existing verification frameworks such as CommonRoad verification framework.

**Research Questions:**
- **RQ1:** What should the architecture of a pipeline look like that enables an LLM-assisted tool to generate executable rules and functions for rule-based verification of map transformations?
- **RQ2:** Which prompting strategies and LLM approaches are most effective for jointly generating logical formulas and corresponding function definitions in this context?
- **RQ3:** How can the effectiveness of the proposed approach be demonstrated through case studies, and which evaluation metrics best capture its utility in real-world transformation verification scenarios?

As a proof of concept, we build on the CommonRoad Scenario Designer [5], a tool capable of transforming maps between formats such as OpenDRIVE and CommonRoad. While the tool supports various geometric and semantic conversions, it currently lacks functionality for handling elevation data. At the same time, a computational FOL-based verification pipeline for transformation correctness already exists [7]. In this work, we extend the verification domain to include elevation and develop an LLM-assisted prototype tool pipeline capable of jointly generating formulas and functions for this new domain, which we evaluate in a targeted case study.

The contributions of this paper are threefold.
1) **Conceptual foundation for tool support in computational FOL verification:** We propose an architecture for an LLM-assisted tool pipeline that jointly generates logical formulas and their corresponding function definitions, thereby answering *RQ1* on how such a pipeline should be structured.
2) **Tool-oriented case study with CommonRoad:** We extend the CommonRoad Scenario Designer [7] and its verification pipeline to include elevation as a new semantic domain. This demonstrates the effectiveness of prompting strategies and LLM-based approaches for jointly producing rules and functions, directly engaging with *RQ2*.
3) **Prototype tool implementation and evaluation:** We implement a semi-automated, human-in-the-loop prototype tool and evaluate it on synthetic scenarios with systematically injected errors. This evaluation demonstrates feasibility and provides evidence toward *RQ3*, concerning how the proposed approach can be assessed in practice.

## II. RELATED WORK

### A. Rule-Based Map Checking and Verification

Several works apply rule-based verification to vector-based HD maps. [13], [14] define checks from national specifications (e.g., topology consistency, format compliance), but these remain checklist-style validations tied to local standards. [15] detects per-element changes in lanelets or crosswalks, yet such element-wise detection lacks the abstraction needed for reasoning about higher-level semantics. [16] validates logical and structural integrity across formats, but relies on predefined templates that limit flexibility. By contrast, the CommonRoad verifier [7] applies computational FOL for composable, explainable rules, offering more general and rigorous reasoning. However, it still requires manual domain-specific predicates, limiting scalability.

### B. Natural Language to Formal Logic with Language Models

Recent work has applied LLMs to generate formal logic from natural language, typically in FOL or MTL. Examples include ChatRule [17] (FOL via prompting), TR2MTL [18] (traffic rules to MTL with chain-of-thought), SafePlan [19] (robotic safety as LTL invariants), and GenAI [20] (OCL from structured requirements). Other recent efforts include nl2spec [21], which translates natural language into LTL with interactive user feedback, and AutoSpec [22], which synthesises program specifications (pre/postconditions, invariants) via LLMs combined with static analysis.

While these approaches show the potential of symbolic rule generation, they all assume that predicates are predefined and fixed. This rigidity limits adaptability to new domains, where both formulas and executable functions are required. Our work bridges this gap by combining logic generation with automatic synthesis of predicate functions, enabling scalable extension of computational FOL verification frameworks.

## III. TOOL APPROACH

### A. Pipeline Overview

As illustrated in Fig. 1, this design directly addresses *RQ1* by defining the architecture of an LLM-assisted tool pipeline that integrates rule generation and function synthesis into the CommonRoad verification framework. The overall workflow consists of two components:
1) **CommonRoad verification framework (existing system):** Contains formal rules, the ANTLR-based rule

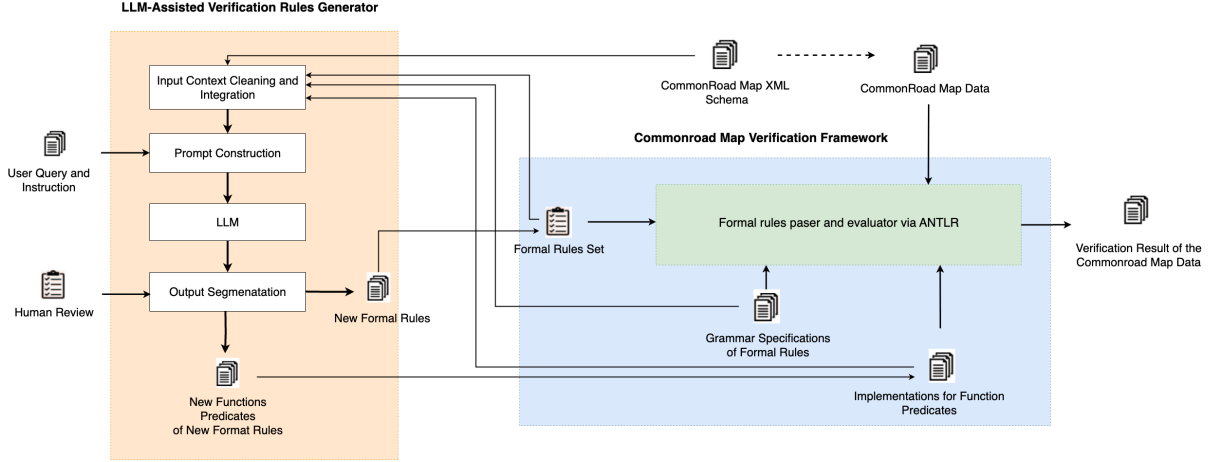

Fig. 1. Architecture of the LLM-assisted verification rules generator based on CommonRoad verification framework [7].

parser and evaluation engine, and existing predicate implementations in Python.

2) **LLM-assisted verification rules generator:** Uses structured context (grammar, rules, predicates, XML schema) to generate new rules and predicates that integrate directly into the framework. Outputs are designed for grammar compliance and executability, and are reviewed by humans before integration to ensure correctness.

### B. Prompt Construction

To generate executable rules and corresponding predicates, we guide the LLM with structured context and strict output constraints. Each prompt includes: (i) context (cleaned inputs; comments removed), (ii) instructional constraints (roles, insertion boundaries), and (iii) a rule specification reflecting domain semantics. This structure addresses *RQ2*, defining strategies and constraints that steer the LLM to produce grammar-compliant rules and executable predicates, with an output contract requiring the LLM to return:

1) a grammar-compliant formal rule,
2) a Python-implementable function predicate (name and signature),
3) a brief semantic explanation.

more details will be introduced in Section IV. This ensures outputs are executable, integrable, and auditable by human reviewers. Potential issues such as syntactically correct but semantically invalid predicates are explicitly handled during the human review stage before integration into the framework.

### C. Rule and Predicate Synthesis

To extend computational FOL to new semantic domains, the LLM jointly generates symbolic FOL formulas and corresponding executable predicates defined by a name, Python implementation, and semantic description. These predicates are then automatically registered in the framework's handler, replacing manual insertion and ensuring consistent expansion of the function library.

## IV. CASE STUDY AND RESULTS

To evaluate our approach, we applied the proposed pipeline to generate elevation-related verification rules, focusing on a domain not previously addressed by existing rule sets. The full case study examples and code are available online[1].

**Experiment Setup.** We designed three representative categories of elevation-related defects involving scenarios like bridges and slopes: (1) *Excessive slope*, where road segments exhibit unrealistic grade values; (2) *Abrupt elevation step*, i.e., sudden height discontinuities between consecutive lanelets; and (3) *Insufficient clearance in overlapping segments*, where two roads overlap in the $z$-dimension but the vertical distance between them is below a safe threshold.

We constructed 40 synthetic scenarios with RoadRunner [23]: 30 scenarios contained up to two of the above defects and 10 scenarios served as defect-free baselines. All scenarios were exported in OpenDRIVE and converted into CommonRoad with elevation, covering typical elevation-related issues.

**Prompt-based Rule Generation.** GPT-4o was employed in a prompt-driven configuration to synthesise formal rules and corresponding function predicates tailored to the above three defect types. To ensure integration feasibility, the user query explicitly instructed the model including above request also as follows: *"Do not modify the ANTLR grammar. Output must comply with the existing syntax and directly integrate with the verification pipeline."*

**Provided Context Artifacts:** The ANTLR grammar definitions, existing predicate implementations, the CommonRoad XML schema and the current rule collection are provided as part of the prompt.

**Generated Outputs.** The pipeline produced formal rules and new predicates aligned with the three defect categories, for example:

- `Is_grade_within_limit(l) || l in L`

---

[1] https://github.com/ZombiesYard/CRUCIBLE

These predicates enforce, respectively, reasonable elevation differences, slope bounds, and minimum clearance in overlapping segments.

**Generated Rule Integration.** Each output underwent human review to validate both its logical correctness and its Python implementation. Once approved, the formal rules were added to the rule collection, and the predicates were implemented and registered in the verification system.

**Results.** The effectiveness of the proposed pipeline was evaluated using both qualitative and quantitative criteria.

*Qualitative assessment:* The pipeline eliminates manual grammar analysis, constraint translation, and function implementation (time-consuming and error-prone), yields syntactically correct and semantically consistent rules, and integrates seamlessly (parsed, linked, executed) without post-processing.

*Quantitative assessment:* Across 40 synthetic maps with three mentioned elevation-defect categories, the framework detected all defect types and produced no false positives on defect-free maps, indicating high precision and robustness.

This evaluation directly targets *RQ3*, combining qualitative and quantitative metrics to demonstrate effectiveness in realistic transformation scenarios.

## V. Conclusion and Future Work

In revisiting our research questions, RQ1 was addressed by designing the tool architecture, RQ2 by developing and testing prompting strategies for rule–predicate synthesis, and RQ3 by evaluating the approach in a controlled case study. These results confirm feasibility and establish a basis for extension.

We presented a semi-automated, human-in-the-loop prototype that leverages LLMs to generate executable verification rules for map transformations, demonstrated with elevation constraints in CommonRoad. The prototype integrates without grammar or framework changes, reduces manual effort, and preserves correctness under human review.

These synthetic experiments provide an initial step but cannot capture the diversity of real-world HD maps. Thus, our evaluation should be viewed as a feasibility study, with future validation on larger and more heterogeneous datasets. Beyond map transformations, the joint generation of formulas and predicates can generalize to domains such as software requirements and cyber-physical system safety. Scaling to broader rule sets and richer logics remains future work.

A key limitation is reliance on GPT-4o, a proprietary model, which raises concerns about reproducibility and semantic errors. We plan to investigate open-source LLMs and automated consistency checks to strengthen robustness and reproducibility.